Research Article

# The Total Solar Irradiance, UV Emission and Magnetic Flux during the Last Solar Cycle Minimum

**E. E. Benevolenskaya[1,2] and I. G. Kostuchenko[3]**

[1] *Pulkovo Astronomical Observatory, Pulkovskoe sh. 65, Saint Petersburg 196140, Russia*
[2] *Saint Petersburg State University, Saint Petersburg 198504, Russia*
[3] *Karpov Institute of Physical Chemistry, Ul. Vorontsovo Pole 10, Moscow 105064, Russia*

Correspondence should be addressed to E. E. Benevolenskaya; benevolenskayae@mail.ru

We have analyzed the total solar irradiance (TSI) and the spectral solar irradiance as ultraviolet emission (UV) in the wavelength range 115–180 nm, observed with the instruments TIM and SOLSTICE within the framework of SORCE (the solar radiation and climate experiment) during the long solar minimum between the 23rd and 24th cycles. The wavelet analysis reveals an increase in the magnetic flux in the latitudinal zone of the sunspot activity, accompanied with an increase in the TSI and UV on the surface rotation timescales of solar activity complexes. In-phase coherent structures between the midlatitude magnetic flux and TSI/UV appear when the long-lived complexes of the solar activity are present. These complexes, which are related to long-lived sources of magnetic fields under the photosphere, are maintained by magnetic fluxes reappearing in the same longitudinal regions. During the deep solar minimum (the period of the absence of sunspots), a coherent structure has been found, in which the phase between the integrated midlatitude magnetic flux is ahead of the total solar irradiance on the timescales of the surface rotation.

## 1. Introduction

The minimum of the solar activity separating the activity cycles 23 and 24 is often called "an unusual minimum." For comparison, during the previous minimum, the lowest annual sunspot number was 8.6 in 1996. During the latest minimum, the annual sunspot number reached lower values: 7.5, 2.5, and 3.1 in 2007, 2008, and 2009, respectively. A typical minimum lasts for about 486 spotless days (http://spaceweather.com/), but since 2004, as many as 821 spotless days were observed. Woods [1] studied the irradiance during the latest solar cycle minimum and compared it with the previous minimum in 1996. He found that the solar irradiance was lower during the latest minimum. Thus, the total solar irradiance (TSI) was smaller in 2008 than that in 1996 by about 200 ppm. The irradiance measured with the SOHO Solar EUV Monitor (SEM) at 26 to 34 nm was by about 15% lower in 2008 than that in 1996. This EUV decrease could be explained by the abundance of low-latitude coronal holes during the latest cycle minimum.

It is known that variations of the sunspot activity during a solar cycle occur because of the generation of the magnetic flux in the convective zone. The magnetic flux is produced by dynamo processes and moves to the solar surface on account of magnetic buoyancy. The emerged magnetic flux impacts the solar luminosity, causing its cyclic behavior. However, the relationship between the magnetic flux and solar irradiance is rather complicated. Various magnetic features such as sunspots, flares, network elements, faculae, and prominences contribute to TSI in different ways. The major contribution to TSI is made by the photosphere. On the other hand, lines formed in the chromosphere and in the corona also impact the total solar irradiance. The development of the sunspot activity is accompanied with total irradiance reduction. This happens only because sunspots on the solar disk in the visible wavelength are dark. And the sunspots, in turn, are dark due to the suppression of the convection in the presence of strong magnetic fields. It is the so-called TSI blocking effect, which makes it difficult to determine the relationship between the magnetic flux and the total solar irradiance.

Fortunately, bright plages or faculae surrounding sunspots contribute to the total solar irradiance, which varies along with the solar cycle.

In a certain way, the brightness of magnetic elements is a function of wavelength [2]. If sunspots display negative fluctuations in the visible wavelengths, they are typically bright in the UV and EUV. It is a consequence of the fact that the complex of solar activity affects all layers of the solar atmosphere. Above the complex of solar activity, there exists a loop structure with a closed configuration of the magnetic field strength, filled by hot plasma emitting in UV, EUV, and X-ray wavelengths.

The irradiance associated with the network elements displays even more complicated behavior. According to the HINODE observations [3], weak magnetic fields enhance the brightness of the quiet sun. However, inner network fields, which are traced by weak horizontal fields, do not indicate the brightness effect. We can suggest, therefore, that the impact of magnetic fields on brightness in quiet regions is mostly due to strong, vertical, small-scale (subpixel) fields.

In this paper, we study the impact of the magnetic flux of the sunspot activity on the solar irradiance, with the purpose to understand the role of the long-lived complexes of the solar activity. In this aspect, the long and deep solar minimum gives a unique opportunity to analyze in detail variations of the total solar irradiance (TSI) and the spectral solar irradiance as a function of the magnetic flux and sunspot area. We can estimate the contribution of each complex of the solar activity to the TSI and UV, separately. We can also compare these results with those for the time when the sunspots do not contribute to the TSI at all.

In addition, the distribution of the sources of the TSI and UV is particularly important for the development of empirical and semiempirical irradiance models [4], for prediction of the solar luminosity.

## 2. Data Analysis

Here, we use the data for TSI and UV (115–180 nm) obtained by the SORCE instruments (http://lasp.colorado.edu/sorce/). The sunspot areas are taken from the webpage (http://solarscience.msfc.nasa.gov/greenwch.shtml). The magnetic data are represented by synoptic maps of the Wilcox Solar Observatory (WSO, http://wso.stanford.edu/).

The solar radiation and climate experiment (SORCE) is a NASA-sponsored satellite mission that is providing state-of-the-art measurements of the incoming X-ray, ultraviolet, visible, near-infrared, and total solar radiation. The SORCE spacecraft was launched on 25 January, 2003. SORCE carries four instruments, including the spectral irradiance monitor (SIM), solar stellar irradiance comparison experiment (SOLSTICE), total irradiance monitor (TIM), and the XUV photometer system (XPS). In our study, we used the daily data of TIM and SOLSTICE from 10 March, 2007, to 23 January, 2010. TIM measures the total solar irradiance (TSI) with an estimated absolute accuracy of 350 ppm (0.035%). Relative variations in the solar irradiance were measured with the accuracy less than 0.001%/yr [5]. The SOLSTICE measurements provide coverage from 115 nm to 320 nm with the spectral resolution of 1 nm, the absolute accuracy better than 5%, and the relative accuracy of 0.5% per year [6].

The Wilcox Solar observatory at Stanford provides us with large-scale low-resolution synoptic maps of the line-of-sight component of the magnetic field strength $B_{||}$ (measured in microTesla, $\mu$T). The resolution of these maps is 5° of Carrington longitude (73 points from 0° to 360°). Each synoptic map consists of 30 data points in equal steps along the latitude sine from −0.97 to 0.97. Using these magnetic data, we calculated the magnetic flux as the absolute values of $B_{||}$. The magnetic flux ($F_{\text{mag}}$) was integrated over a box with a width of ±90° in the longitude and with a height of ±40° in the latitude. $F_{\text{mag}}$ was calculated as a function of Julian date, moving the box along the Carrington longitude. This procedure makes it possible to take into account the evolution of the magnetic flux ($F_{\text{mag}}$) from east to west on the solar limb relatively to the center of the solar disk.

We analyzed the variable fluxes of the total solar irradiance (TSI), ultraviolet emission (UV) (115–180 nm), integrated daily sunspot area, and integrated midlatitudinal magnetic flux ($F_{\text{mag}}$) during the time 10 March 2007 to 23 January 2010. Figure 1 presents all the studied variables as a function of time. Here, time is in Julian dates, in years, and in days since the first day of the time series. Figure 1(a) (upper plot) depicts daily values of the TSI. Figure 1(a) (bottom plot) displays variations of the ultraviolet emission integrated over the wavelength range 115–180 nm. The integrated midlatitude magnetic flux, $F_{\text{mag}}$, is represented in Figure 1(b) (upper plot). Bottom plot (Figure 1(b)) shows the daily integrated sunspot area in thousandths of the solar hemisphere. The daily integrated sunspot area is estimated by summing the area of all sunspots per day.

A simple comparison of the variables TSI and $F_{\text{mag}}$ indicates their consistency, except for the times of strong TSI negative variations, which occurred due to the sunspot blocking effect. The increase in the daily integrated sunspot area corresponds to that in the integrated midlatitude magnetic flux. There is a strong correlation between the $F_{\text{mag}}$ and the intensity of the UV. As we can see in Figure 1, during the deep minimum, the integrated midlatitude magnetic flux corresponds to low luminosity in the total and UV irradiance.

## 3. A Nonaxisymmetric Pattern of the Solar Irradiance

We analyzed the TSI and magnetic data ($B_{||}$) in the form of Carrington maps using the method described in [7]. First, the time series (TSI) is interpolated with the reference to the beginning ($to_i$) and the end ($tn_i$) of each Carrington rotation "$i$" with the step ($tn_i − to_i$)/27. The synodic period of the Carrington rotation is 27.2753 days. Thus, the first day of the interpolated time series corresponds to 360° of Carrington longitude for Carrington rotation CR2055, and the last day of this time series corresponds to 0° of Carrington longitude CR2091. TSI is represented in the form of 2D matrix, one dimension of which is the Carrington longitude with the step 360°/27 and the second is the time measured in the numbers of Carrington rotations (CR2055 to CR2091). This approach makes it possible to study features with the lifetime longer than one Carrington period. Figure 2 presents the 2D distribution of TSI for CR2055–CR2091 (b). The stacked WSO synoptic maps of the line-of-sight component of the magnetic field strength during the studied time are shown in Figure 2(a). This figure reveals the nonuniform longitudinal distribution of the magnetic flux and TSI related to the occurrence and development of solar activity complexes. The complexes of the solar activity are marked by "A," "B," and "C." The solar complexes "B" and "C" are related to the strong fluctuations of the sunspot area within the defined longitudinal zone about 180°–270°.

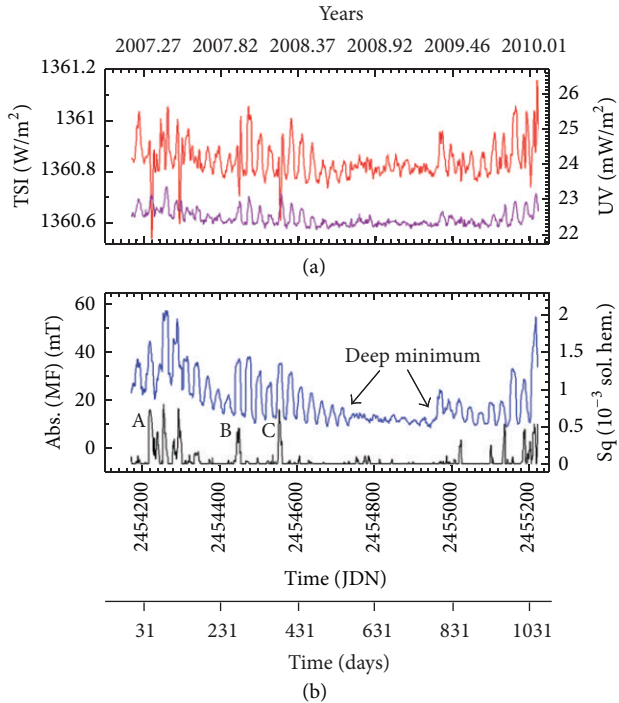

Figure 1: The daily values of the total irradiance (TSI) during the time 10 March 2007–23 January 2010 ((a), upper red line) in W/m$^2$; the variations of the UV in the range of 115–180 nm ((a), bottom purple line) in mW/m$^2$; the WSO absolute values of the magnetic field strength of the line-of-sight component (magnetic flux, $F_{mag}$) integrated over the latitude from −40° to 40° and from −90° to 90° over the longitude ((b), upper blue line) in milliTesla (mT); the daily integrated sunspot area is represented as thousandths of the solar hemisphere ((b), bottom black lines). The beginning and the end of the deep minimum are marked by arrows.

The solar complex "A" corresponds to the multiple strong variations of the sunspot area (Figure 1(b), bottom plot), which occur during several Carrington rotations and are spread along the longitude. But only one of these fluctuations, the one related to the complexes of solar activity, exists inside the aforementioned longitudinal zone.

It is clearly seen that, during CR2055–2076 (March 2007–November 2008), a strong magnetic activity exists in the longitudinal zone 200°–300° and rotates with the velocity rate slightly exceeding that of the Carrington rotation. This magnetic structure, combined from "A," "B," and "C" activity complexes, contributes to the TSI due to the surrounding bright faculae. The longitudinal zones in which a strong solar activity occurs during several Carrington rotations are called active longitudinal zones. They are related to a source of the magnetic field under the photosphere.

It is interesting that a long-lived longitudinal activity occurred at the ascending phase of the cycle 24 during the Carrington rotation CR2107–CR2115 at the longitude 300°–360° and latitude 13°–22° [8]. This complex produced a strong X-flare (X6.9) in September, 2011. It was the first long-lived complex in the cycle 24. At the beginning of the solar cycle 23, a similar longitudinal pattern existed from June, 1996, to June, 1998 [7]. During this time, the TSI distribution displayed an increase at longitudes 200°–300°. The EIT/SOHO data for the extreme UV irradiance indicated the increase in the coronal temperature associated with the activity complex at the beginning of the cycle 23. Therefore, the heating of the solar corona is closely related to long-lived complexes of the solar activity and to a corresponding source of the magnetic field under the photosphere.

The complexes of the solar activity "A," "B," and "C" are seen during CR2055–CR2063 (31 March, 2007–1 December, 2007), CR2064–CR2067 (1 December, 2007–20 March, 2008), and CR2068–CR2072 (20 March, 2008–3 August, 2008), respectively (Figure 2). They are maintained by the magnetic flux emerging from a subsurface source. The complex "A" consists of five fluctuations of the daily summarized area (Figure 1(b), bottom plot), but only three of them are strong. They exist mainly due to long-lived sunspots. Moreover, the blocking effect is related only to the 1st and the 3rd strong fluctuations of the sunspot area. What sunspots are responsible for these negative TSI fluctuations?

During the time of evolution of the complex "A," five long-lived sunspots were observed. The first long-lived large sunspot in the region 0953 NOAA appeared on 26 April, 2007 (S14° E73°, CL308°), and lasted for 14 days. Here, CL indicates the Carrington longitude. This region reached its maximum area of about 520 millionths of the solar hemisphere on 29 April, 2007 (according to daily sunspot summaries from the NOAA Space Environment Center). The second sunspot 0956 NOAA began to be observed on 15 May, 2007 (N02° E61°, CL070°), and disappeared on 24 May, 2007. On 18 May, 2007, its maximum area was about 300 millionths of the solar hemisphere. The third sunspot 0960 NOAA appeared on 2 June, 2007 (S06°, CL181°), and went behind the solar limb on June, 14. Its area reached about 540 millionths of the solar hemisphere on 4 June, 2007. The fourth region 0961 NOAA appeared on 26 June, 2007 (S09° E77°, CL219°), and moved behind the solar limb on 8 July, 2007. On 1 July, 2007, the maximum area was about 210 millionths of the solar hemisphere. The last, fifth region 0963 NOAA appeared on 8 July, 2007 (S09° E83°, CL054°), and moved behind the solar limb on 20 July, 2007 (its maximum area was about 530 millionths of the solar hemisphere on 11 July, 2007). The three strongest fluctuations of the sunspot area that belong to the complex "A" (Figure 1(b), bottom plot) correspond to the long-lived sunspots 0953 NOAA, 0960 NOAA, and 0963 NOAA. However, the strong blocking effect of the total solar irradiance occurs only due to the regions 0953 NOAA and 0963 NOAA. Why the long-lived sunspot 0960 NOAA with the area of about 540 millionths of the solar hemisphere does not block the total solar irradiance is still unclear.

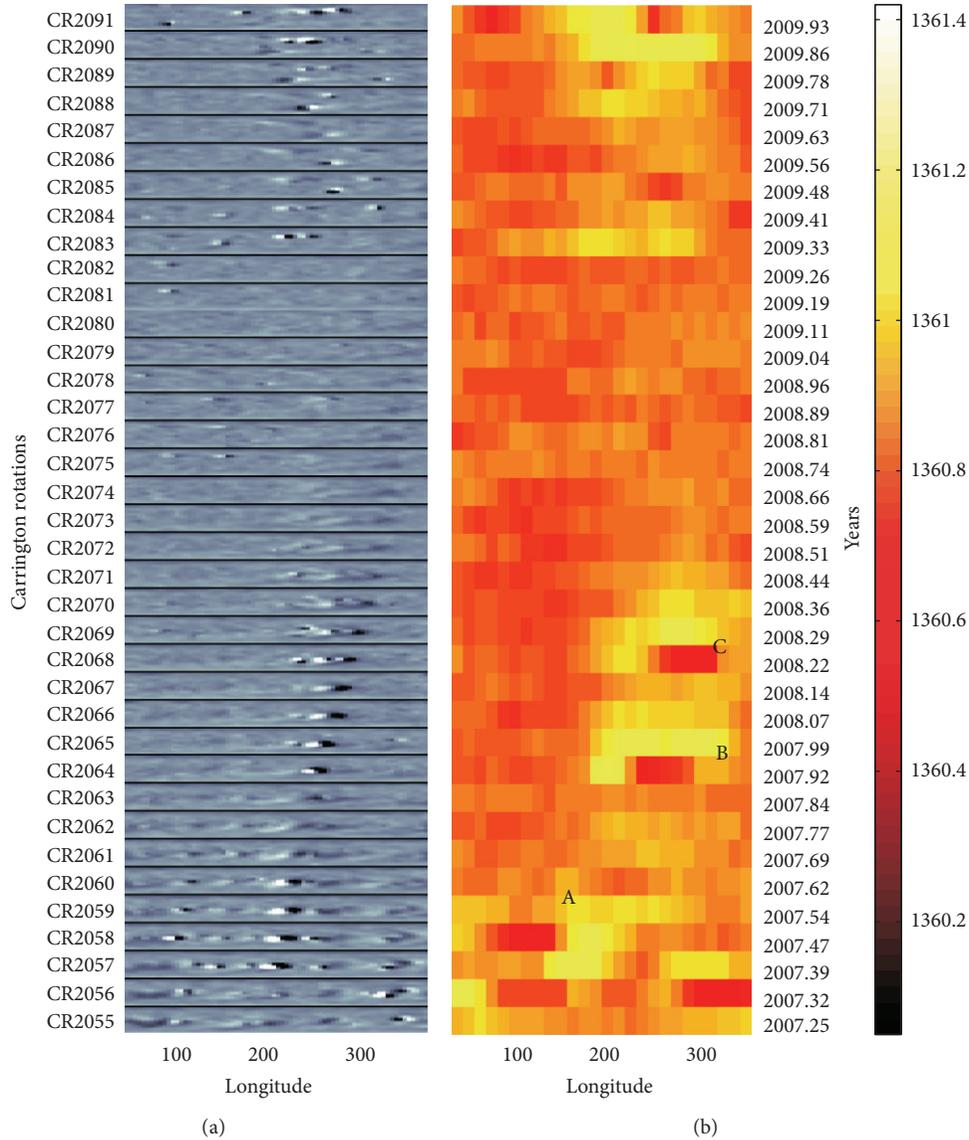

Figure 2: (a) Stacked WSO synoptic maps, 10 March, 2007, to 23 January, 2010, in gray scale from −250 to 250 microTesla; (b) TSI as a function of the Carrington number and the Carrington longitude. Complexes of solar activity are marked with A, B, and C. The time scale on the right indicates the beginnings of Carrington rotations. The color bar shows the TSI intensity in W/m$^2$.

The activity of B and C complexes is of special importance, as they belong to the overlapping cycles 23 and 24. According to the Hale's law, the polarity of the preceding and following parts in bipolar complexes of the solar activity alters from one solar cycle to the next. However, within the time interval of the so-called "cycles' overlap," the activity complexes of both polarities coexist.

During 2008-2009, the "old" magnetic flux (which belonged to the cycle 23) was concentrated in longitudinal zones, and the largest part of the "new" flux (which belonged to the cycle 24) with reversed magnetic polarity emerged in the same zones, within the longitude interval 180°–270° [9]. The complex marked by "B" appeared in the region 0978 NOAA and lasted from 7 to 19 December, 2007. The sunspot 0981 NOAA (N30°, CL246°) emerged in the same longitudinal zone as the sunspot 0980 NOAA with the "old" polarity (S06°, CL239°). The complex marked by "C" began since 28 March, 2008, when the cycle 23 returned and three big sunspots (0987 NOAA, 0988 NOAA, and 0989 NOAA) appeared. Their magnetic flux displayed the polarity of the "old" solar cycle 23. After that, the sun was practically blank, without sunspots (see Figure 2(a)).

Generally, at the beginning of a "new" cycle, sunspots with the "new" polarity appear at latitudes 25° to 35°, while those with the "old" polarity, located close to the equator, disappear. However, as we mentioned, there is a time of the cycles' overlap, when sunspots with both "old" and "new" polarities coexist. On 5 October, 2008, the sunspot 1003 NOAA of the new cycle appeared in the southern hemisphere at 23° of latitude and at 222° of Carrington longitude. It disappeared rapidly, and during the following several days we observed plages instead of the sunspot in the same location. On 11 October, 2008, a small sunspot (1004 NOAA) emerged at the latitude S08° and Carrington longitude 188° and another (1005 NOAA) at a higher latitude (N26°, CL116°). During the following day, two plages, 1003 NOAA (S23°, CL222°) and 1004 NOAA (S08°, CL188°), coexisted with the northern sunspot 1005 NOAA with the coordinates N26° and CL116°. In CR2075-CR2076 (September-October, 2008), "new" magnetic flux spread over longitude, but it was weak and short lived. Therefore, during the years 2008-2009, the sunspot activity was low, and magnetic regions with the "old" and "new" polarity coexisted.

To study the relationship between the nonuniform distribution of the solar magnetic flux, TSI, and UV, we applied the wavelet analysis.

## 4. Wavelet Analysis of TSI, UV, and $F_{\text{mag}}$

We used MATLAB package to perform cross-wavelet and wavelet coherence analysis developed by Grinsted et al. [10]. This software includes a code originally written by C. Torrence and G. Compo (available at http://paos.colorado.edu/research/wavelets/) and by E. Breitenberger from the University of Alaska, which was adapted from the SSA-MTM Toolkit freeware (http://www.atmos.ucla.edu/tcd/ssa/).

Wavelet analysis solves problems by decomposing a time series into time and frequency spaces simultaneously. We obtain the information on both the amplitude of any "periodic" signal within the series and the variation of this amplitude with the time. The wavelet transformation is generally used to analyze time series that contain nonstationary power at many different frequencies. Grinsted and his colleagues [10] recommend using the Morlet wavelet with $\omega_0 = 6$, since it provides a good balance between time and frequency localization. In this case, the Morlet wavelet scale is almost equal to Fourier period. According to the description presented in [10], Morlet wavelet which consists of a plane wave modulated by a Gaussian is defined as

$$\psi_0(\eta) = \pi^{-1/4} \exp^{i\omega_0 \eta} \exp^{-\eta^2/2}, \quad (1)$$

where $\omega_0$ is the dimensionless frequency, $s$ is the scale, and $\eta = st$. The continuous wavelet transformation of a discrete sequence $X_n$ is defined as the convolution of $X_n$ with a scaled and normalized wavelet:

$$W_n^X(s) = \sqrt{\frac{\delta t}{s}} \sum_{n'=0}^{N-1} x_{n'} \psi^* \left[ \frac{(n'-n)\delta t}{s} \right], \quad (2)$$

where "∗" indicates the complex conjugation. Finally, the wavelet power spectrum is defined as $|W_n(s)|^2$.

Figure 3 presents the Morlet continuous wavelet power spectrum for three sets of normalized time series (magnetic flux, ultraviolet emission, and total solar irradiance). Here, we indicate the cone of influence to display the statistical significance of the wavelet power (marked by solid lines in Figure 3). The continuous wavelet transformation displays edge artifacts, because the wavelet is not completely localized in time. It is therefore useful to introduce a cone of influence (COI), in which the edge effects cannot be ignored. Here, we take COI as the area in which the wavelet power caused by a discontinuity at the edge drops with the factor of $e^2$ [10].

The Morlet continuous wavelet transformation reveals common features with some differences in three time series. In the case of $F_{\text{mag}}$ (Figure 3(a)), significant peaks occur on the timescale of the solar activity rotation during the descending phase of the solar cycle 23 and at the beginning of the solar cycle 24. The wavelet power shows separation between the complexes "A" and "B&C" within the same time intervals (Figure 3(b)). The separation increases for the wavelet power of TSI. Significant peaks for "A" are expanded to smaller time intervals of about 10 days. The TSI and UV series also display high values of the power spectrum in the band with the maximum at about 120–130 days during 200–500 days. This time interval coincides with the lifetime of the solar activity complexes "B" and "C."

## 5. Cross-Wavelet Transformation and Coherence Structure of TSI, UV, and $F_{\text{mag}}$

If a cross-wavelet power transformation reveals areas with high common power spectrum in the time-frequency space, then the wavelet coherence transformation finds local phase-locked behavior in this space. The cross-wavelet transformation (XWT) of two series $X$ and $Y$ can be defined as $W_n^{XY}(s) = W_n^X(s) W_n^{Y*}(s)$, where "∗" denotes complex conjugation. The wavelet transformations of the series $X$ and $Y$ are $W_n^X(s)$ and $W_n^Y(s)$, where $s$ is the scale, so that $\eta = s \cdot t$; $\eta$ is the dimensionless time. The cross-wavelet spectrum is complex and can be defined as the cross-wavelet power spectrum $|W_n^{XY}(s)|$. Another useful parameter derived from the wavelet analysis is the wavelet transformation coherence (WTC) defined as the square of the cross-spectrum (XWT) normalized to the individual power spectrum. Phase coherence is defined as $\tan^{-1}[\text{Im}[|W_n^{XY}(s)|]/\text{Re}[|W_n^{XY}(s)|]]$ [10].

Figure 4(a) presents the cross-wavelet power spectrum of the TSI and $F_{\text{mag}}$. This figure indicates strong correlation between the irradiance and the magnetic flux during the descending phase of the cycle 23 and during the overlap of cycles 23 an 24 (up to 500th day of the time series) with the maximum at periods close to that of the solar activity surface rotation. The relative phase relationship of the two data series is shown with arrows. Arrows pointing to the right show the in-phase behavior of both selected data sets in the time-frequency space; those pointing to the left indicate the antiphase behavior of the series. Both time series are in-phase during the evolution of the complexes of solar activity "B" and "C" (arrows pointing to the right). But there is no in-phase behavior of the TSI and $F_{mag}$ in the presence of the complex activity marked by "A." Note that this complex consists of several activity complexes separated in the Carrington longitude. Then, during the deep minimum (500–800 days of the time series), the cross-wavelet power spectrum of the TSI and $F_{mag}$ does not reveal strong common areas in the time-frequency space (Figure 4(a)). Further, significant peaks appear again due to the development of the solar cycle 24.

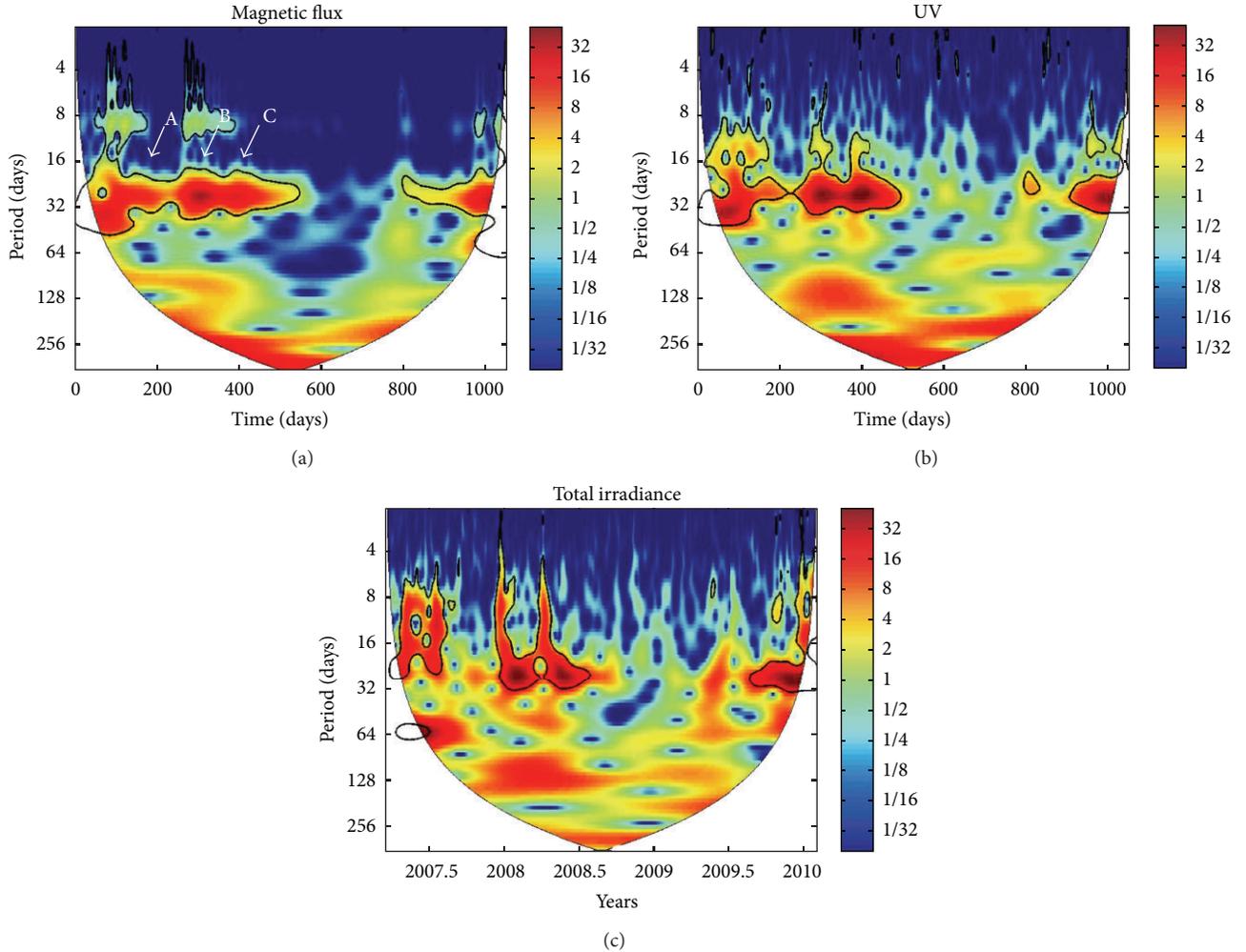

FIGURE 3: Morlet continuous power spectrum of the magnetic flux (a), ultraviolet emission (b), and total irradiance (c). The thick black contour designates the 5% significance level against the red noise. The cone of influence (COI), where the edge effects might distort the picture, is not shown by color. Complexes of solar activity contributing to TSI and UV are marked by A, B, and C in (a).

On the periods of about 120 days, cross-wavelet transformation of the TSI and $F_{mag}$ does not show in-phase behavior of the power spectrum in the time-frequency space. This may be related to the fact that strong magnetic flux coincides with dark sunspots. An increase in the magnetic flux of the sunspots results in the reduction of the TSI.

The cross-wavelet transformation power spectrum for UV and $F_{mag}$ is slightly different (Figure 4(c)). The cross-wavelet of the UV and magnetic flux indicates the in-phase behavior for all three complexes of solar activity on the rotational timescale (the synodic period is about 27-28 days).

Moreover, we observe the in-phase behavior of the cross-wavelet transformation power spectrum between the UV and $F_{mag}$ on timescales of about 100–140 days.

The wavelet transformation coherence (WTC) makes it possible to find structures in the time-frequency space, which display locally phase-locked behavior. Such a coherence structure indicates that variations of both time series are in the same phase in the local time-frequency space.

In Figures 4(b) and 4(d), the arrows pointing to the right indicate an in-phase behavior of TSI and UV, respectively. Arrows pointing to the left indicate the antiphase behavior; those pointing downward show that the first series of data is by 90° second; those pointing upward indicate that the second data series is by 90° ahead of the first.

The wavelet transformation coherence of the TSI and $F_{mag}$ and that of the UV and $F_{mag}$ display a detailed picture of coherent structures in time-frequency space. There is a significant difference between the TSI, $F_{mag}$ series, on the one hand, and the UV, $F_{mag}$ series, on the other hand, in the coherence areas on periods of the surface rotation. The UV and magnetic flux coherence structures associated with the magnetic activity (Figure 4(d)) are in-phase (arrows point to the right). The stronger magnetic flux causes stronger UV emission. Also, we observe the extended coherence structure on time intervals from about 20 to 50 days during the descending phase of the cycle 23. The coherence structure related to the beginning of the cycle 24 appears slightly earlier then the development of sunspot activity begins (see Figures 4(d) and 1(b)).

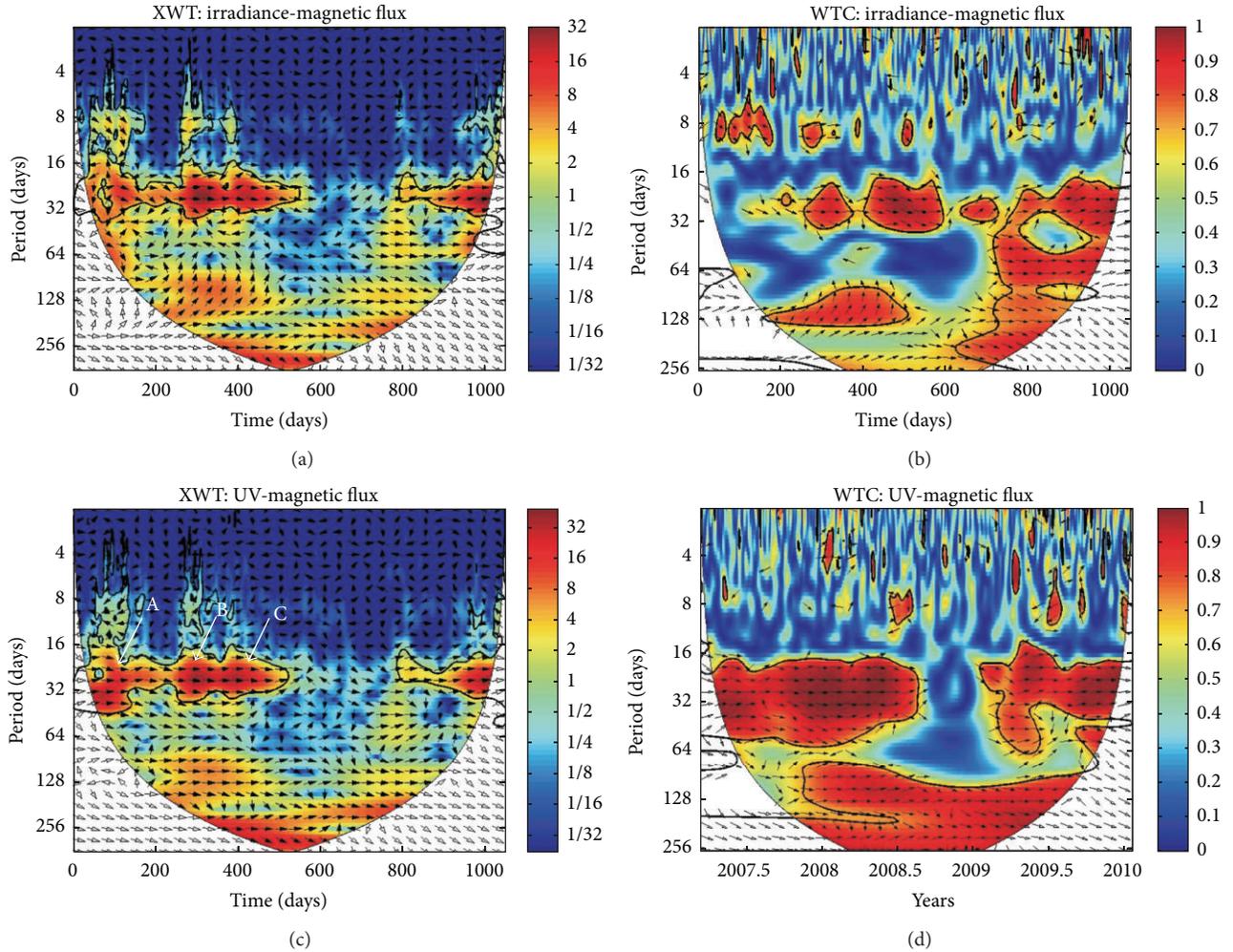

FIGURE 4: (a) Cross-wavelet transformation of the irradiance and the magnetic flux; (b) coherence of the irradiance and the magnetic flux; (c) cross-wavelet transformation of the ultraviolet and the magnetic flux; (d) coherence of the ultraviolet and the magnetic flux. Arrows indicate the phase relationship between the two data series in time-frequency space: (1) arrows pointing to the right show the in-phase behavior; (2) arrows pointing to the left indicate the antiphase behavior; (3) arrows pointing downward show that the first data series is by 90° ahead of the second series; (4) arrows pointing upward indicate that the second data series is by 90° ahead of the first series. Complexes of solar activity contributing to the TSI and UV are marked by A, B, and C in (c).

The coherence structure of the TSI and the $F_{mag}$ on periods of the surface activity rotation is decomposed into several structures during the time of overlapping of cycles 23 and 24 (Figure 4(b)) and during the solar cycle minimum. Thus, complex "A" does not create a coherent structure, but "B" and "C" display two coherent regions on periods of the surface activity rotation. The separation is related to the sunspot blocking effect. The second coherent region in frequency-time space shifts to the region of the deep minimum. The third coherent structure during the deep minimum shows that the phase of the $F_{mag}$ is ahead of the total solar irradiance. This effect requires explanation. It may be a result of the influence of the polar irradiance (due to bright polar faculae), which is in the antiphase with the magnetic flux at midlatitude during the solar minimum. In the case of the UV, the relationship between the polar irradiance and the midlatitude magnetic flux is in-phase due to the coronal holes visible in X-ray, UV, and EUV emissions. On the periods of about 120 days, we see the coherent structure, in which the phase of the $F_{mag}$ is ahead of the TSI phase.

Therefore, long-lived complexes of the solar activity are maintained by the magnetic flux, which appears in the same longitudinal zones. These emerged fluxes create the in-phase coherent structure of the magnetic flux and the total solar irradiance due to bright faculae. The complex "A," consisted of regions, which spread over the Carrington longitude, does not produce a strong coherent structure on the surface rotational timescales. However, the complexes "B" and "C" display strong coherence structures on surface rotational timescales.

## 6. Discussion

The total solar irradiance (TSI) describes the total radiant energy, in the form of electromagnetic radiation emitted by the sun at all wavelengths, that falls for each second on 1 square meter outside the Earth's atmosphere, a value proportional to the "solar constant" introduced earlier in the last century. TSI is a measure of the solar energy flux. The radiative flux decreases when dark sunspots are present on the disk and due to bright faculae or plages [11].

Fligge and Solanki [12], using a reconstruction of the total solar irradiance since 1700, found that long-term TSI variations exist due to the evolution of the solar network or other processes. Therefore, connections between the solar magnetic activity and total irradiance are based on the following factors: irradiance reductions that result from the passage of dark sunspots across the solar disk; irradiance increases due to facular brightening; and long-term irradiance variations related to changes in the solar network. Moreover, the long-term and solar cycle variations are obliged to the existence of long-lived complexes of solar activity with extended bright plages or faculae. There is a good consistency between the TSI, UV, and the irradiance reconstructed from Ca K faculae area [13]. Usually, a large and long-lived complex of the solar activity is surrounded by a large area of faculae. Moreover, sometimes a part of an emerging bipolar magnetic flux is represented by bright faculae rather than by dark sunspots in continuum. Bright and expended plages can explain the brightening in 2D maps of TSI (Figure 2(b)).

Long-lived longitudinal nonuniformity in the solar activity (active longitudes) has been known for a long time, but generally, this phenomenon is related to active phases of solar cycles, when there are a number of (several) sunspot complexes and solar magnetic field has a complicated structure. Here, we observed a pronounced and stable longitudinal nonuniformity for the relatively weak solar magnetic flux during the long solar minimum and two cycles overlapping.

## 7. Conclusions

A pronounced longitudinal nonuniform emerging of faculae and sunspots and, as a result, a nonaxisymmetrical distribution of the total and spectral solar irradiance were observed during the long minimum between cycles 23 and 24.

During the descending phase of the solar cycle 23, sunspots reappeared in the same plages, in a limited longitudinal zone, during 22 Carrington rotations. The Carrington longitude of these spots and faculae consecutively shifted from 180° to 270°, demonstrating the rotation rate slightly greater than the Carrington's. After the long minimum, sunspots of the cycle 24 tended to appear in the same longitudinal zone, while the spots of the "old" cycle still exist.

The revealed tendencies make it possible to suppose that a long-lived local source of the emerging magnetic flux exists under solar photosphere. Our results also suggest that conditions for such a source to appear are stable relative to the process of cycle changing. According to the solar dynamo theory, the solar magnetic field during the minimum is expected to be a pure dipole. However, the existence of such a local long-lived subsurface source of magnetic field apparently indicates generation of a nonaxisymmetrical component of the solar magnetic field due to dynamo process.

Our comparative and cross-wavelet analyses show that the increase in the magnetic flux in the latitudinal zone of the sunspot activity is accompanied by the increase in the TSI and UV on the timescale of the rotation of the solar activity complexes. The coherent structures between the midlatitude magnetic flux and TSI/UV occurred in the case of the existence of the long-lived complexes of the solar activity. Therefore, the nonuniform longitudinal distribution of the long-lived solar magnetic activity affects the solar irradiance. Indeed, the found coherent structures are associated with the development of such complexes of the solar activity. Moreover, coherent structures between the TSI and the magnetic flux confirm the idea about the interrelation of the activity processes going on the sun in total.

A similar longitudinal pattern was observed at the beginning of the solar cycle 23 from June 1996 to June 1998 [7]. During this time, the longitudinal distribution of the total solar irradiance, EUV irradiance from the transition region and corona, and solar magnetic flux integrated over solar disk increase in the same longitudinal zone (200°–300°). We concluded that precisely the long-living complexes of sunspot activity associated with this longitudinal zone make a significant contribution to the variation of the total solar irradiance and to the heating of the solar corona.

Therefore, the revealed nonaxisymmetrical character of the solar magnetic field turned out to be quite stable during the solar minimum, and the solar irradiance is closely related to the nature of the solar magnetic field.

## Acknowledgments

The authors thank Dr. Grinsted, Professor Moore, Dr. Jevrejeva, Dr. Torrence, Dr. Combo, and Dr. Breitenberger for the development of the Wavelet MATLAB package. Also, they are thankful to the science team of SORCE and WSO for making their data available for free download. Part of this work is supported by the Program 22 of the Russian Academy of Science.

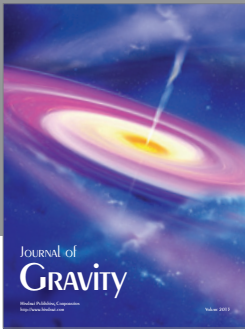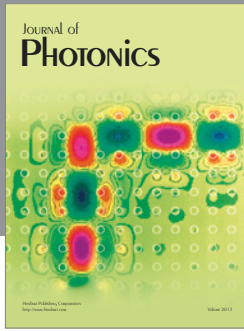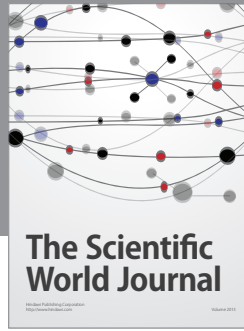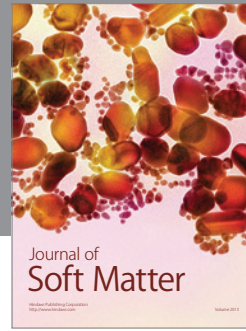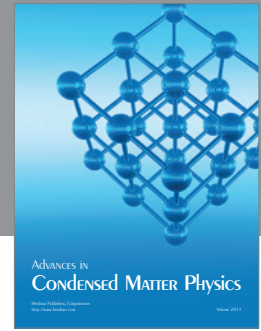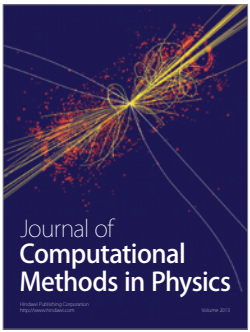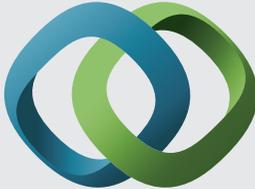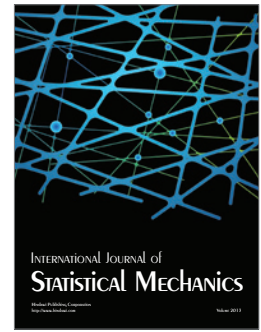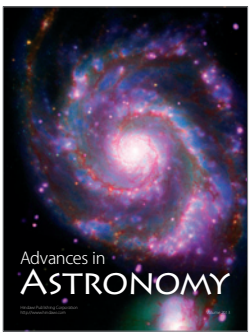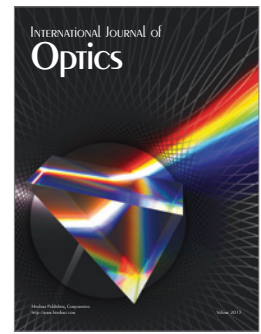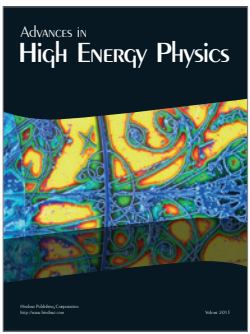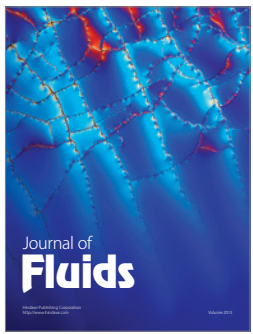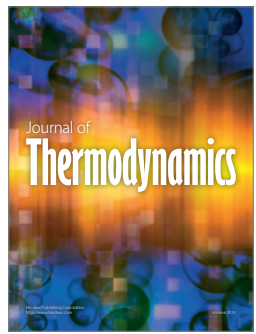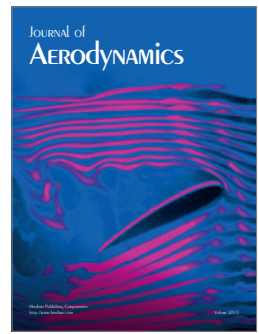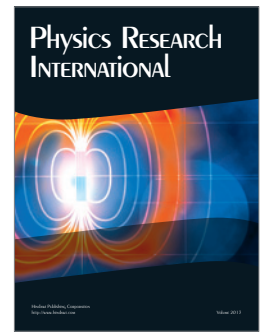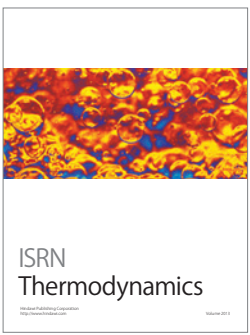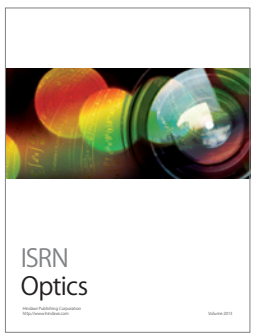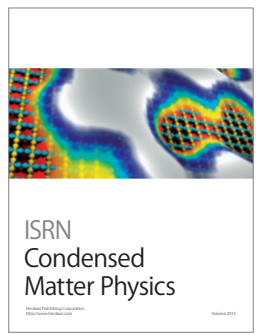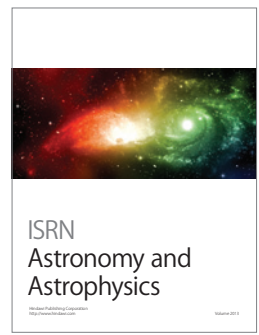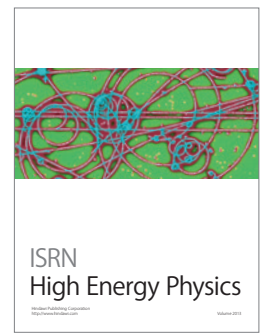